\documentclass[12pt,a4paper]{article}
\usepackage[dvips]{graphicx}
\usepackage{amsmath}
\usepackage{subfigure}
\usepackage{cite}
\usepackage{a4wide}

\begin{document}

\hfill \today

\begin{center}

\vspace{32pt}

  { \bf The Structure of Projected Center Vortices in Lattice Gauge 
Theory}

\end{center}

\vspace{18pt}

\begin{center}
{\sl R. Bertle${}^a$, M. Faber${}^a$, J. Greensite${}^{b,c}$,
and {\v S}. Olejn\'{\i}k${}^d$}

\end{center}

\vspace{18pt}

\begin{tabbing}

{}~~~~~~~~~~~\= blah  \kill
\> ${}^a$ Inst. f\"ur Kernphysik, Technische Universit\"at Wien, \\
\> ~~A--1040 Vienna, Austria.  E-mail: {\tt faber@kph.tuwien.ac.at} \\
\\
\> ${}^b$ Physics and Astronomy Dept., San Francisco State Univ.,\\
\> ~~San Francisco, CA~94117, USA. E-mail: {\tt greensit@stars.sfsu.edu}
\\
\\
\> ${}^c$ Theory Group, Lawrence Berkeley National Laboratory,\\
\> ~~Berkeley, CA~94720, USA. E-mail: {\tt greensit@lbl.gov}\\
\\
\> ${}^d$ Institute of Physics, Slovak Academy of Sciences, \\
\> ~~SK--842 28 Bratislava, Slovakia.  E-mail: {\tt fyziolej@savba.sk}

\end{tabbing}

\vspace{18pt}

\begin{center}

{\bf Abstract}

\end{center}

\bigskip

We investigate the structure of center vortices in maximal center
gauge of SU(2) lattice gauge theory at zero and finite temperature. In
center projection the vortices (called P-vortices) form connected two
dimensional surfaces on the dual four-dimensional lattice. At zero
temperature we find, in agreement with the area law behaviour of
Wilson loops, that most of the P-vortex plaquettes are parts of a
single huge vortex. Small P-vortices, and short-range fluctuations of
the large vortex surface, do not contribute to the string tension. All
of the huge vortices detected in several thousand field configurations
turn out to be unorientable. We determine the Euler characteristic of
these surfaces and find that they have a very irregular structure with
many handles. At finite temperature P-vortices exist also in the
deconfined phase. They form cylindric objects which extend in time
direction. After removal of unimportant short range fluctuations they
consist only of space-space plaquettes, which is in accordance with
the perimeter law behaviour of timelike Wilson loops, and the area law
behaviour of spatial Wilson loops in this phase.

\vfill

\newpage

\section{Introduction}

It is very well known that the center symmetry is of crucial importance 
for
QCD on the lattice. At the beginning of eighties it was shown that the 
deconfinement phase transition is connected with the spontaneous 
breaking of
center symmetry~\cite{KPS81}.
In the confined phase field configurations are center symmetric, 
leading to
symmetric distributions of Polyakov loops and to infinite energy of 
single
quarks. In the deconfined phase one center element is favoured by the 
Polyakov
loop distribution, quark charges can be screened by the gluon field.

There is now increasing evidence that center symmetry is not only 
relevant as an order parameter for confinement, but is also the crucial 
concept in understanding how confinement comes about.
The idea that center vortices are responsible for confinement was 
put forward at the end of 70's \cite{tH79,Ma79,Co79,NO79,Fe81}, but
numerical evidence in favor of this idea is rather recent
\cite{PRD97,DJ97,PRD98b,EL98,Forc}.  Our principle tool for locating 
vortices,
and investigating their effect on gauge-invariant quantities,
is ``center projection'' in maximal center gauge. 
Maximal center gauge is a gauge where all link variables are 
rotated as close as possible to center elements of the gauge group. 
Center projection is a mapping from the SU(2) link variables to 
$Z_2$ link variables.  ``P-vortices'', formed from plaquettes
with link product equal to $-1$ (we will call them P-plaquettes), 
are simply the center vortices of the projected $Z_2$ gauge-field
configurations.

   P-vortices on the projected lattice are thin, surface-like objects,
while center vortices in the unprojected configuration should be 
surface-like
objects of some finite thickness.  Our numerical evidence indicates 
that
P-vortices on the projected lattice locate thick center 
vortices on the unprojected lattice.  We also find that these thick
vortices are physical objects, and that the
disordering effect of such vortices is responsible for the entire QCD
string tension.  Details may be found in ref.~\cite{PRD98b}; a 
discussion
of Casimir scaling in the context of the the center vortex theory
is found in ref.~\cite{PRD98a} and \cite{APS99}.

In this article we discuss the structure of P-vortices in 
center-projected
field configurations.  In section two we show that P-vortices tend to 
form
very large vortices (as required if they are the driving mechanism 
behind quark confinement, see also ref.\ \cite{Po98}), and that
small-scale fluctuations of the vortex surfaces 
don't contribute to the string-tension.  
 Further, we determine the 
orientability and the Euler characteristic of these vortices. In 
section three
we investigate the structure of P-vortices at finite temperature.
We find, in agreement with Langfeld et al.~\cite{LT98}, that in
the deconfined phase the vortices are oriented along timelike
surfaces in the dual lattice, and are closed by the lattice periodicity;
thereby explaining the differing behaviour of timelike and spacelike
Wilson loops in this phase.

\clearpage

\section{P-vortices at zero temperature}

\subsection{Finding P-vortices}

  P-vortices are identified by first fixing to 
the direct version of maximal 
center gauge, which in SU(2) gauge theory maximizes
\begin{equation}
\sum_{x,\mu} \left | \text{ Tr} U_\mu(x) \right |^2 \;.
\end{equation}
Then we map the SU(2) link variables $U_\mu(x)$ to $Z_2$ elements
\begin{equation}
Z_\mu(x) \; = \; \text{sign Tr} [ U_\mu(x) ] \;.
\end{equation}
The plaquettes with $Z_{\mu,\nu}(x) = Z_\mu(x) Z_\nu(x+\hat{\mu}) 
Z_\mu(x+\hat{\nu}) Z_\nu(x) = -1$ are the ``P-plaquettes.'' 
The corresponding dual plaquettes, on the dual lattice,
form the closed surfaces (in D=4 dimensions) associated with P-vortices.
This can be easily understood by constructing such a P-vortex out of a 
trivial
field configuration with links $Z_{\mu}=1$. By flipping a single link 
to
$Z_{\mu}=-1$ the six space-time plaquettes attached to this link form 
an
elementary P-vortex. In dual space the corresponding plaquettes form 
the
surface of a cube. By flipping neighbouring links in real space we get 
cubes
in dual space attached to each other. The dual vortex is the connected 
surface
of these cubes.

  The distribution of these P-vortices in space-time determines 
the string tension.  The 
number $n$ of piercings of P-vortices through a Wilson loop determines 
the
value of the projected Wilson loop $W_{cp}(I,J)$ of size $A = I \times 
J$
\begin{equation}
W_{cp}(I,J) \; = \; (-1)^n \;.
\end{equation}
If $p$ is the probability that a plaquette belongs to a P-vortex, then 
by
assuming the independence of piercings we get for the expectation value 
of
$W_{cp}(I,J)$
\begin{equation}
\langle W_{cp}(I,J) \rangle \; = \; \left [ (1-p)1 \; + \; p(-1) \right 
]^A \;
= \; (1-2p)^A \; = \; e^{- \sigma_{cp} A} \; \approx \; e^{-2pA} \;,
\end{equation}
where the string tension in center projection is
\begin{equation}\label{sigmatop} 
\sigma_{cp} \; = \; - \text{ln} (1-2p) \; \approx \; 2p \;.
\end{equation}
\begin{figure}[ht]
\centerline{\scalebox{0.9}{\includegraphics{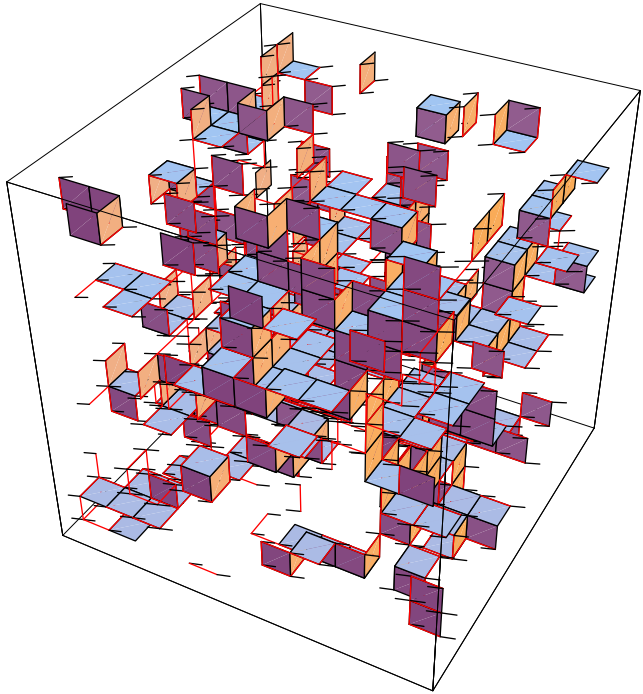}}}
\caption{3-dimensional cut through the dual of a $12^4$-lattice at 
$\beta=2.3$. P-plaquettes are indicated. They form closed 
two-dimensional
surfaces. P-plaquettes which extend in time direction and connect the 
given
3-dimensional space with neighbouring time-slices are indicated by 
small
amputated lines in forward or backward direction.}
\label{math23}
\end{figure}
However, for small vortices the independence assumption of piercings is 
not fulfilled, simply because one 
piercing is always correlated with another piercing nearby, and 
therefore
small vortices in the average do not contribute to the area law 
fall-off of
large Wilson loops. For this reason, confining vortex configurations 
will
have to be very large, with an extension comparable to the size of the
lattice.  Fig.~\ref{math23} shows the P-vortex 
plaquettes in a single time-slice of an equilibrium field configuration 
on a
$12^4$-lattice at $\beta=2.3$.

\begin{figure}
\centerline{\scalebox{0.8}{\includegraphics{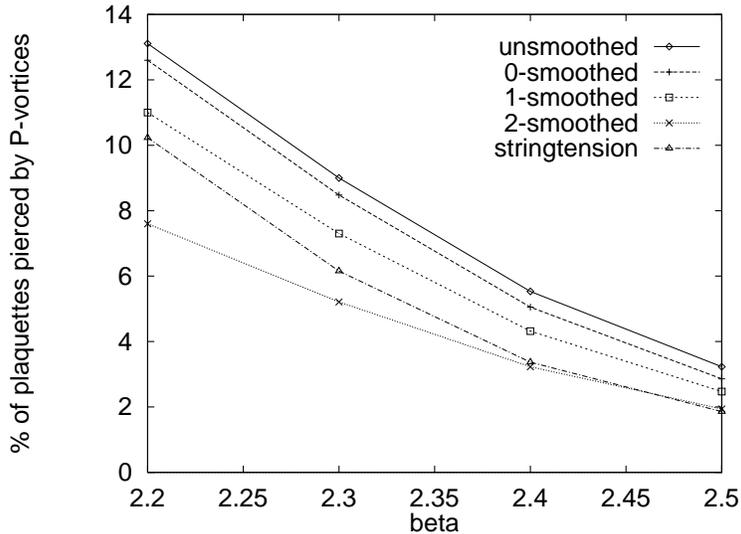}}}
\caption{Percentage $p$ of P-vortex plaquettes for various 
$\beta$-values and
smoothing steps is compared with the fraction $f$ of plaquettes which 
should
be pierced according to the asymptotic string tension $\sigma$.}
\label{nplnegav}
\end{figure}

The above defined probability $p$ scales nicely with the inverse 
coupling
$\beta$. This was shown in refs.~\cite{PRD98b,EL98}. However, already 
from
Fig.~1 of ref.~\cite{PRD97} it can be seen that the $\chi(1,1)$ Creutz 
ratios
lie above the asymptotic string tension $\sigma$, and therefore the 
values of
$p$ come out higher than those of $f$ that can be inferred from 
$\sigma$ using
\begin{equation}\label{sigmatof}
f \; = \; (1-e^{-\sigma})/2\;.
\end{equation}
The values of $p$ and $f$ are shown in Fig.~\ref{nplnegav} for various 
$\beta$-values. Here $p$ is denoted as ``unsmoothed'' (to stress the 
difference from values extracted using a smoothing procedure that will 
be
introduced in section 2.2), while $f$ is denoted as ``stringtension'' 
in
Fig.~\ref{nplnegav} and was extracted from $\chi(3,3)$ Creutz ratios of 
the
center projected field configurations. As shown in ref.~\cite{PRD98a} 
the
values of $\chi(3,3)$ are in good agreement with the determinations of 
the
string tension by Bali et al.~\cite{BS95}. Other data sets shown in 
Fig.~\ref{nplnegav} denoted as ``$n$-smoothing'' will be discussed in 
section
2.2.

There appears the interesting question about the difference between $p$ 
depicted in Fig.~\ref{nplnegav} under the name ``unsmoothed'' and $f$. 
Introducing various smoothing methods of P-vortices we will show below 
that
this difference originates in short distance fluctuations of P-vortex 
surfaces.

\subsection{Size of P-vortices}

One of the most basic properties of vortices which is connected with 
the above
raised questions is their size. In order to check the size and number 
of
P-vortices we determine neighbouring P-plaquettes, which can be easily 
done on
the dual lattice. In most cases there is no doubt about the 
neighbouring
plaquettes, since most dual links connect only two dual P-plaquettes. 
But in
some cases there appear ambiguities, when dual links are attached to 4 
or 6
dual P-plaquettes. Below we investigate these ambiguities in more 
detail
and discuss possible resolutions.

In general P-vortices can be in contact at sites, links and plaquettes. 
Let's
have a look at these different possibilities. A contact point is of no 
importance since we define the connectedness via common links. For our 
case of
the $Z_2$ center group a plaquette belonging to two independent 
vortices leads
to a fusion of these vortices. In this context, it is important to 
understand
if the irregular structure of P-vortices depicted in Fig.~\ref{math23} 
is due to such fusion process of simpler vortices. The remaining
possibility, a contact of P-vortices at links needs a more detailed 
discussion.

\begin{figure}[ht]
\centerline{\scalebox{0.85}{\includegraphics{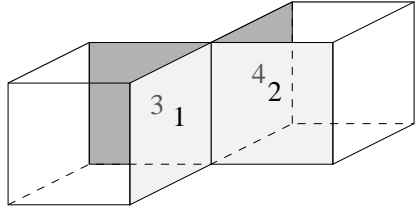}}}
\caption{The question of connectedness can not be solved uniquely for 
this configuration of P-vortices.}
\label{cubtouch}
\end{figure}

A simple example of such field configurations are two cubes touching 
like in
Fig.~\ref{cubtouch}.
For such a configuration in dual space it is not clear whether it 
builds one
or two vortices. At the given length scale there is no unique solution 
for the
question of connectedness. Connecting 1 with 2, 3 with 4 would result 
in one
vortex, connecting 1 with 3, 2 with 4 in two separated vortices. In 
most
cases the situation can be resolved by postponing the decision about 
the
connectedness of these plaquettes until, by following the vortex 
surface
in all other directions, the indicated plaquettes (usually) turn out 
to be members of the same vortex. There appear some cases where no 
decision is possible by these means, as
in the simple example of Fig.~\ref{cubtouch}. In order to get a lower 
limit
for the size of vortices we decide in such cases to treat the 
configuration as
two separate vortices.
\begin{figure}[b!]
\centerline{\scalebox{0.8}{\includegraphics{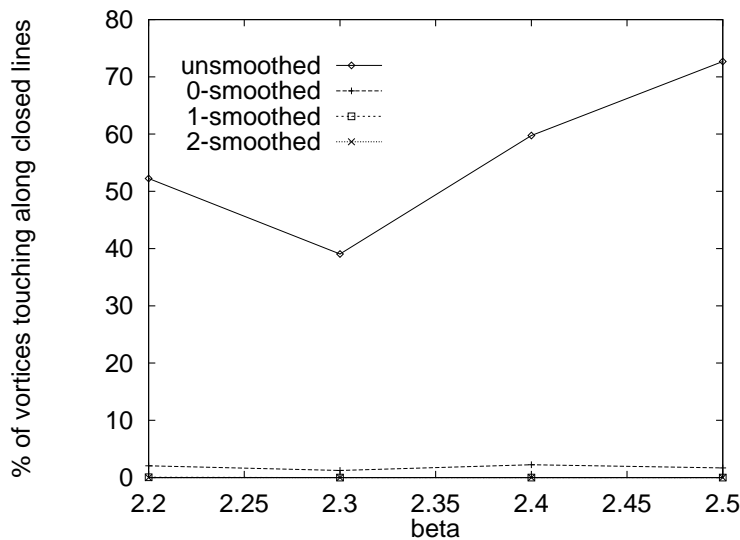}}}
\caption{Percentage of such P-vortices which touch along closed lines 
for
various types of smoothing.}
\label{NDCL}
\end{figure}

It may even occur that vortices touch along closed lines. In these 
cases parts
of the vortex surface can't be reached following regular connections of 
plaquettes. These cases are even not so rare, their percentage is shown 
in
Fig.~\ref{NDCL}. The length of the closed line is usually very small 
and
includes in the average 5 to 7 links.

With the above mentioned rules for deciding connectivity of P-vortices
in ambiguous cases, we determine the P-vortex sizes. Since most of the 
plaquettes turn out to 
belong to the same vortex the most interesting vortex is the largest.
\begin{figure}
\centerline{\scalebox{0.8}{\includegraphics{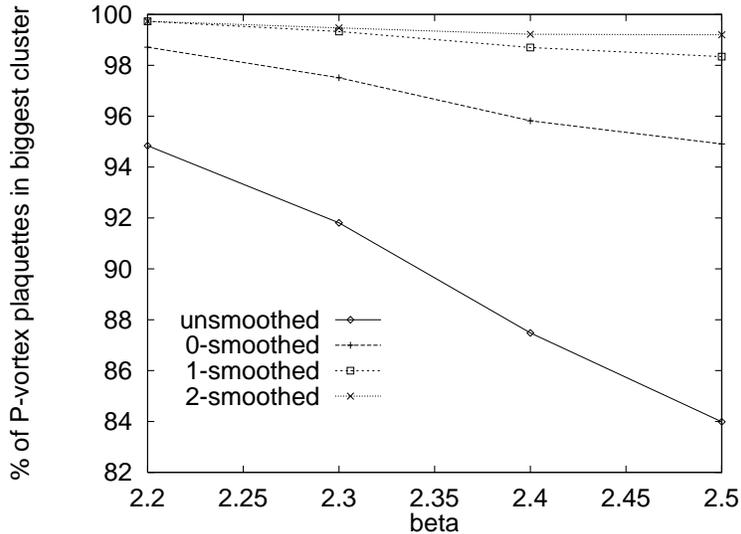}}}
\caption{Percentage of P-plaquettes in the largest vortex for various 
types
of smoothing.}
\label{BGCL}
\end{figure}
The full line in Fig.~\ref{BGCL} shows the percentage of P-vortices 
belonging
to the largest vortex for various values of $\beta$, see also 
\cite{Be98}.
For  the evaluation at beta=2.2, 2.3, 2.4 and 2.5 we used 2000, 2000, 
800 and
240 field-configurations on $12^4$-, $12^4$-, $16^4$-  and 
$22^4$-lattices,
resp. It is obvious that for $T=0$ and all investigated $\beta$-values 
there
is mainly one huge vortex, which contains 
around 90\% of all P-plaquettes. All 
other vortices are rather small and should not contribute to the string 
tension according to the above given arguments. Small vortices means 
strongly
correlated piercings resulting in a perimeter contribution to Wilson 
loops.
P-vortices of diameter $d$ lead to correlations for distances larger 
than $d$.
Only vortices extending over the whole space contribute to an area law 
at all
length scales. We conclude that the string tension is determined by the 
area
of a single huge P-vortex. 

   The fact that P-vortices in the confined phase have an extension 
comparable to the lattice size is a recent result of Chernodub et al.\ 
\cite{Po98}, and
our finding is consistent with theirs.  The fact that there appears
to be only \emph{one} very large vortex may also be related to results 
reported
by Hart and Teper \cite{Hart}.  These authors find that large monopole 
loops,
identified in abelian projection, intersect to form one huge cluster.
Since our previous studies \cite{DJ97} indicate that abelian-projection 
monopoles loops occur mainly on P-vortex surfaces, it seems quite
natural that large loops on a single large surface would tend to 
intersect.

\subsection{Small-scale fluctuations of P-vortices}

In the preceding section we gave 
an argument why small vortices do not contribute 
to the string tension. By the same argument, small fluctuations of the 
vortex
surface affect only perimeter law contributions. 
We will remove these short range 
fluctuations from P-vortex surfaces and show that the percentage of 
plaquettes
belonging to such smoothed vortices gives directly the string tension 
$\sigma$.
\begin{figure}
\centering
\centerline{\scalebox{0.35}{\includegraphics{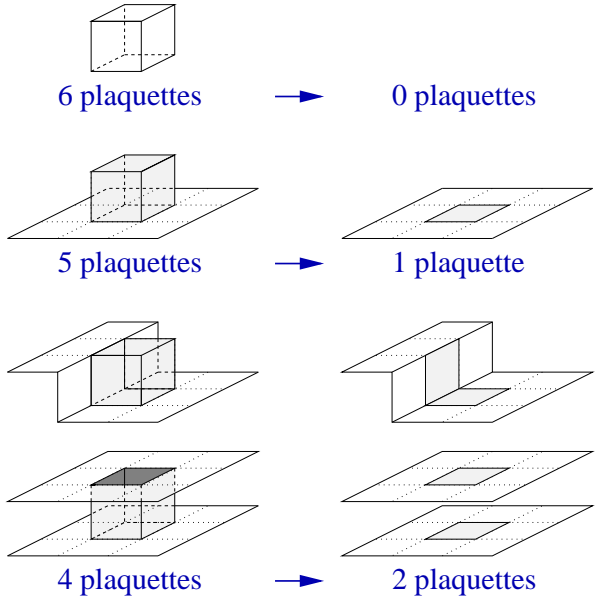}}}
\caption{Various smoothing steps for vortices.}
\label{smoothing}
\end{figure}

\begin{figure}
\centerline{\scalebox{0.9}{\includegraphics{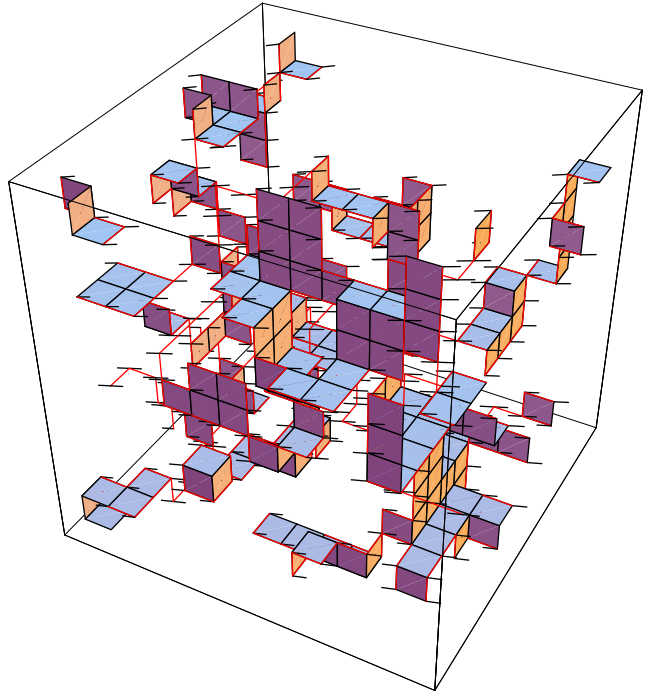}}}
\caption{Result of the configuration of Fig.~\ref{math23} after 
2-smoothing.
P-plaquettes which extend in time direction and connect the given 
3-dimensional space with neighbouring time-slices are indicated by 
small
amputated lines in forward or backward direction.}
\label{math23_4}
\end{figure}

In order to follow this idea we introduce several smoothing steps which 
are
depicted in Fig.~\ref{smoothing}. In a first step we identify single 
isolated
P-vortex cubes consisting of six dual P-plaquettes only and remove 
them. Since
we substitute in this step 6 plaquettes by 0 we call this step 
0-smoothing. In
the next step called 1-smoothing we identify cubes covered by 5 
P-vortex
plaquettes. Such cubes can be substituted by one complementary 
plaquette which
closes the cubes. Finally, we substitute cubes with 4 plaquettes by the 
complementary 2 plaquettes. There are two different arrangements for 
these 4,
resp. 2 plaquettes. In the 2-smoothing step we substitute both of them 
in
accordance with Fig.~\ref{smoothing}. In order to visualize the effect 
of
smoothing on the appearance of vortices we show in Fig.~\ref{math23_4} 
the
result of 2-smoothing for the configuration in Fig.~\ref{math23}.

In Fig.~\ref{BGCL} we compare for various smoothing steps the 
probability that
a P-plaquette belongs to the largest vortex. It is clearly seen that 
the
largest reduction in the number of vortex plaquettes is already 
achieved with
0-smoothing. After 2-smoothing the probability reaches more than 99 \% 
which
shows that only very few small vortices survive the smoothing 
procedure. In
all investigated field configurations we found a single huge P-vortex, 
we never met a configuration with two large vortices.

The relation between the percentage $p_i$ of P-plaquettes after 
$i$-smoothing
and the fraction $f$ which one would expect from the physical string 
tension
$\sigma$ according to Eq.~\ref{sigmatof} can be seen from 
Fig.~\ref{nplnegav}.
One can see that the value of $p_i$ nicely approach $f$, especially for 
the
larger values of $\beta$ where the P-plaquettes get less dense.

\begin{figure}
\centering
\centerline{\scalebox{0.8}{\includegraphics{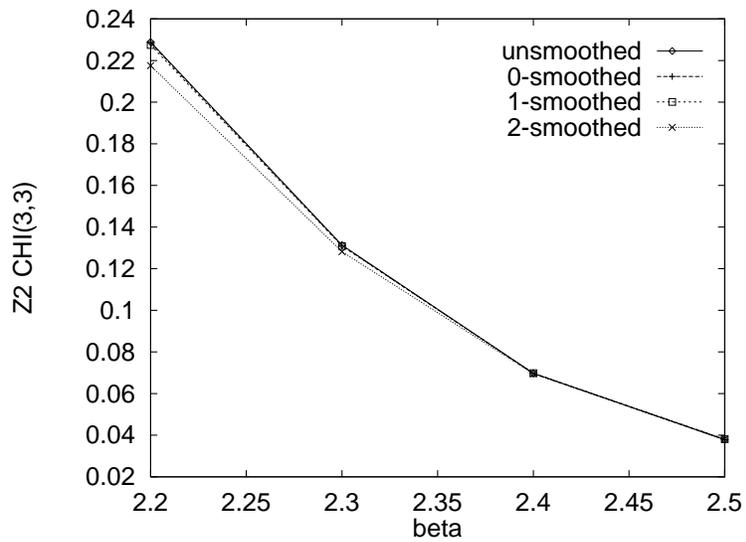}}}
\caption{Creutz ratios for different smoothing steps.}
\label{creutz}
\end{figure}

Further we check the Creutz ratios extracted from P-configurations 
after
various smoothing steps. The results are shown in Fig.~\ref{creutz}.
It is clearly seen that 0- and 1-smoothing does not change the 
extracted
Creutz ratios. At $\beta = 2.2$, where the percentage of P-vortex 
plaquettes is of the order of 10\%, 2-smoothing causes some small ($\sim
5\%$) reduction in $\chi(3,3)$, but even this small deviation goes away
at larger $\beta$.

   Center projected Creutz ratios $\chi(R,R)$ are nearly independent of
$R$.  Only $\chi(1,1)$ (the logarithm of $1 \times 1$ loop) deviates 
from the
Creutz ratios for large loops. This behaviour indicates that the only 
significant correlation among P-plaquettes is at a distance of one
lattice spacing. The short-range fluctuations which may be responsible 
for this correlation are removed by our smoothing procedure. When this 
is
done, the values of $p$ and $f$ defined above come together, and the
string tension is almost unchanged.

\subsection{Topological properties}

The basis for the investigation of the topology of P-vortices is the 
following rule: The type of homomorphy of a surface can be determined 
if it is
connected, compact and closed. Then, it is determined by a) the 
orientation
behaviour, b) the Euler characteristic. P-vortices in dual space would 
fulfill
the requested conditions if every link joined only two plaquettes. As 
mentioned above this is not always the case.

Therefore, we will proceed in the following way. First, we will treat 
the
determination of the orientability for the case that every link joins 
uniquely
two attached P-vortex plaquettes. For each link of every P-plaquette we 
specify a sign. We start with an arbitrary P-plaquette and fix an 
arbitrary
rotational direction. Those two links which are run through in positive 
axis
direction get a plus sign, the other two get a minus sign. We continue 
at an
arbitrary neighbouring P-plaquette. Its rotational sense we fix in such 
a way
that the joining link gets the opposite sign than before. We continue 
this
procedure for every plaquette of the given P-vortex. If at the end 
every link
of the P-vortex has two opposite signs we call such P-vortices 
orientable. The
simplest example is a three-dimensional cube. If some links appear with 
two
equal signs the P-vortex is unorientable, e.g. with the topology of a 
Klein
bottle.

We already gave a certain classification for those cases where four or 
six
P-plaquettes are joined by a single link, see also Fig.~\ref{cubtouch}. 
In
cases of this kind, where different pairs of P-plaquettes can be
treated as belonging to 
independent vortices, we determine the orientability as though the 
vortices did not touch at any link. With this procedure we 
get an upper bound for the orientability of vortices. Analogously we 
proceed
for the case of a vortex touching itself at a link. We determine the 
orientability for a configuration where this touching is avoided.

The simulation shows that without exceptions the large vortices in all 
investigated field configurations, for all investigated $\beta$-values, 
turned
out to be unorientable surfaces. We checked for the various 
employed smoothing steps whether this behaviour remains unchanged. 
It turns out that P-vortices remain unorientable after smoothing;
apparently the smoothing procedure does not remove all of the local 
structures (e.g. ``cross-caps'') responsible for the global 
non-orientability.

The second property which determines the topological properties of 
P-vortices
is the Euler characteristic $\chi$ which is defined by
\begin{equation}
\chi \; = \; {\cal N}_0 \; - \; {\cal N}_1 \; + \; {\cal N}_2 \;,
\end{equation}
where ${\cal N}_k$ is the number of $k$-simplices: ${\cal N}_0$ is the 
number
of vertices, ${\cal N}_1$ the number of links, and
${\cal N}_2$ the number of plaquettes. $\chi$ is directly related 
to the genus $g$ of a surface, in the orientable case by
\begin{equation} \label{Euler-orient}
\chi \; = \; 2 \; - \; 2g
\end{equation}
and in the unorientable case by
\begin{equation} \label{Euler-unorient}
\chi \; = \; 2 \; - \; g \;.
\end{equation}

An orientable surface of genus $g$ is homeomorphic to a sphere with $g$ 
attached handles.  An unorientable surface of genus 
$g$ corresponds to a sphere with $g$ attached M\"obius strips 
(also known as ``cross-caps'').

The determination of the Euler characteristic of a P-vortex is not 
inhibited
by possible self-touchings. We can simply treat the vortex as it is; 
the result is the average between a possible separation and a real 
fusing of the two parts of the vortex.  For a detailed discussion of 
this
case, we refer the interested reader to ref.\ \cite{Be98}.

\begin{figure}
\centering
\centerline{\scalebox{0.8}{\includegraphics{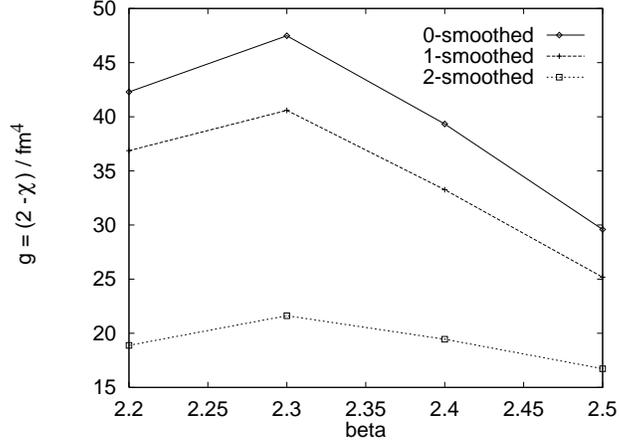}}}
\caption{Genus $g$ of P-vortices per fm$^4$ for various inverse 
couplings
$\beta$.}
\label{euler}
\end{figure}
In Fig.~\ref{euler} we show the genus $g = 2 - \chi$ per fm$^4$ of 
P-vortices for various values
of $\beta$. For the prediction of these values we used 
$\sqrt{\sigma} /\Lambda = 58$ and $\sqrt{\sigma}=440$ MeV. Without 
smoothing the
genus takes a maximal value around $\beta=2.3$. 
With 0-smoothing, only elementary vortices are removed and therefore 
$\chi$ is
unchanged. With 1-smoothing, contact points and contact links can be 
removed,
therefore a reduction of the genus of a vortex by 1-smoothing is to be 
expected. In Fig.~\ref{euler} this reduction is of the order of 15 \%. 
By
2-smoothing also regular bridges in vortices can be removed, decreasing 
in
this way the genus. This reduction amounts to 55\% at $\beta = 2.2 $ 
and 43\%
at $\beta = 2.5$. After 2-smoothing the genus stays more nearly
constant with $\beta$ than without 
smoothing. It has to be investigated how the behaviour of $g$ behaves 
at still
higher $\beta$-values. The trend in the investigated region seems 
compatible with a scaling behaviour for genus $g$, and is not 
compatible
with a self-similar 
short-range structure below the confinement length scale. Fractal
structure of that kind would lead
to an increase of the genus with $\beta$, as more handles are uncovered
at ever-shorter length scales. Of course, even a smoothed P-vortex 
surface will be rough at length scales beyond the confinement scale, 
and an appropriate
fractal dimension can be defined.  The fractal dimension of unsmoothed
P-vortex surfaces, using the definition of dimension $D=1 + 2A/L$, where
$A$ is the number of plaquettes and $L$ the number of links on the 
vortex
surface, has been reported in refs. \cite{Po98} and \cite{BVZ99}.

These investigations of the topology of P-vortices show that they are 
not
topologically 3-spheres. This is not so surprising; there was
no particular reason that vortices \emph{should} have this topology.
The structure which we identified, huge vortices extending over the 
whole
lattice, unorientable with a lot of handles, is quite consistent with 
rotational symmetry in four dimensions. But as we will see, at 
finite temperature this symmetry can be destroyed.

\section{Topology of P-vortices at finite temperature}

   The first discussion of the confinement/deconfinement phase
transition in the context the vortex theory and center-projection
methods, was made by Langfeld et al.\ in ref.\ \cite{LT98}, who give
a nice explanation of the space-space string tension in the
deconfined phase in terms of vortices closed in the time direction
by lattice periodicity.  Another very interesting investigation
into the effect of finite temperature on vortex structure is 
due to Chernodub et al. \cite{Po98}.
In this section we will extend our study of P-vortex topology, and
the effect of our smoothing steps on P-vortices, to the finite 
temperature case. 

We did our finite temperature calculations on a $2\cdot 12^3$-lattice 
for
$\beta$-values between 1.6 and 2.6 and on a $4\cdot 12^3$-lattice for 
$\beta$-values between 1.8 and 2.6. With a heat-bath-algorithm we 
measured
after 1000 equilibration steps 1000 configurations with a distance of 
20 for
each investigated $\beta$-value.

\begin{figure}
\centering
\centerline{\scalebox{0.8}{\includegraphics{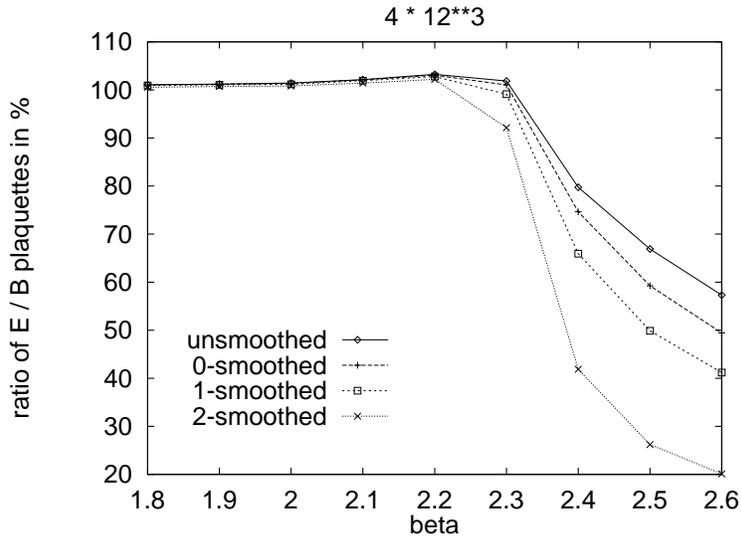}}}
\caption{Ratio of space-time (E) and space-space (B) P-plaquettes as a 
function of $\beta$ for a $4\cdot 12^3$-lattice.}
\label{EB4-asymmetry}
\end{figure}

The most striking difference to zero temperature calculations is the 
strong
asymmetry of P-plaquette distributions in the deconfined phase which 
can be
seen in Fig.~\ref{EB4-asymmetry}.
\begin{figure}
\centering
\centerline{\scalebox{0.8}{\includegraphics{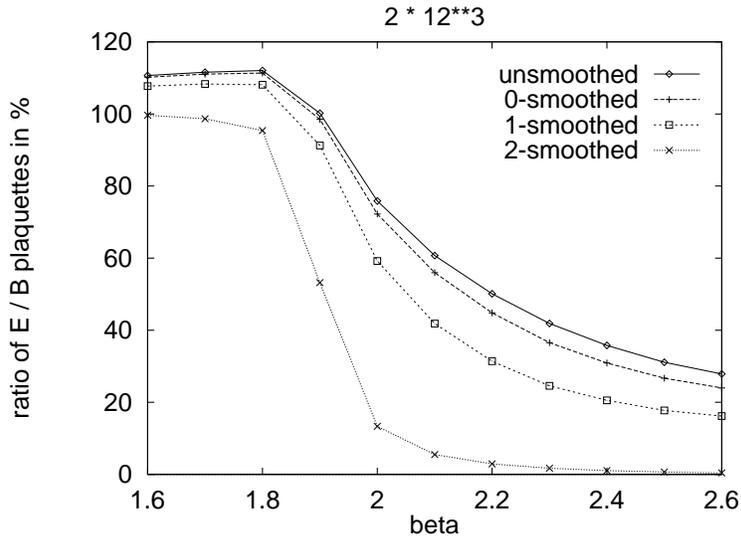}}}
\caption{Same as Fig.~\ref{EB4-asymmetry} for a $2\cdot 12^3$-lattice.}
\label{EB2-asymmetry}
\end{figure}
As a short-notation we use E-plaquette for space-time and B-plaquette 
for
space-space P-plaquettes. An investigation of this asymmetry was also 
performed in \cite{LT98}.  Just below the phase transition the density 
of E-plaquettes is slightly larger than the density of B-plaquettes. 
The
excess is even larger for a time extent of 2
lattice units where it amounts to more than 10 \% as one can see in 
Fig.~\ref{EB2-asymmetry}. The excess at $N_t=2$ seems to be connected 
with short range fluctuations of the vortices, since it is greatly 
reduced
by smoothing. The detected strong asymmetry in the deconfined
phase gives a very intuitive explanation for the behaviour of 
space-time and
space-space Wilson loops, as previously discussed in ref.\ \cite{LT98}. 
The dominant vortex which percolates through the lattice is a (mostly)
timelike surface on the dual lattice, which is closed via periodicity in
the time direction.
Polyakov lines are not affected by timelike vortex surfaces, and 
timelike Wilson loops are also unaffected.  Therefore the string tension
of timelike loops is lost in the deconfinement phase.  On the other 
hand,
large timelike vortex surfaces (composed of B-plaquettes) do disorder
spacelike Wilson loops, which accounts for the string tension of spatial
loops (c.f.\ \cite{Karsch}) in the deconfinement regime.

In the deconfined phase, the density of E-plaquettes is strongly 
decreasing
with smoothing and for 2-smoothing soon reaches values close to 0\% as 
seen
in Figs.~\ref{EB4-asymmetry} and \ref{EB2-asymmetry}. E-plaquettes in 
the
deconfined phase appear obviously due to short range fluctuations and 
can't
contribute to an area law behaviour.

\begin{figure}
\vspace{15mm}
\centering
\subfigure[\hfill]{\scalebox{0.60}{\includegraphics{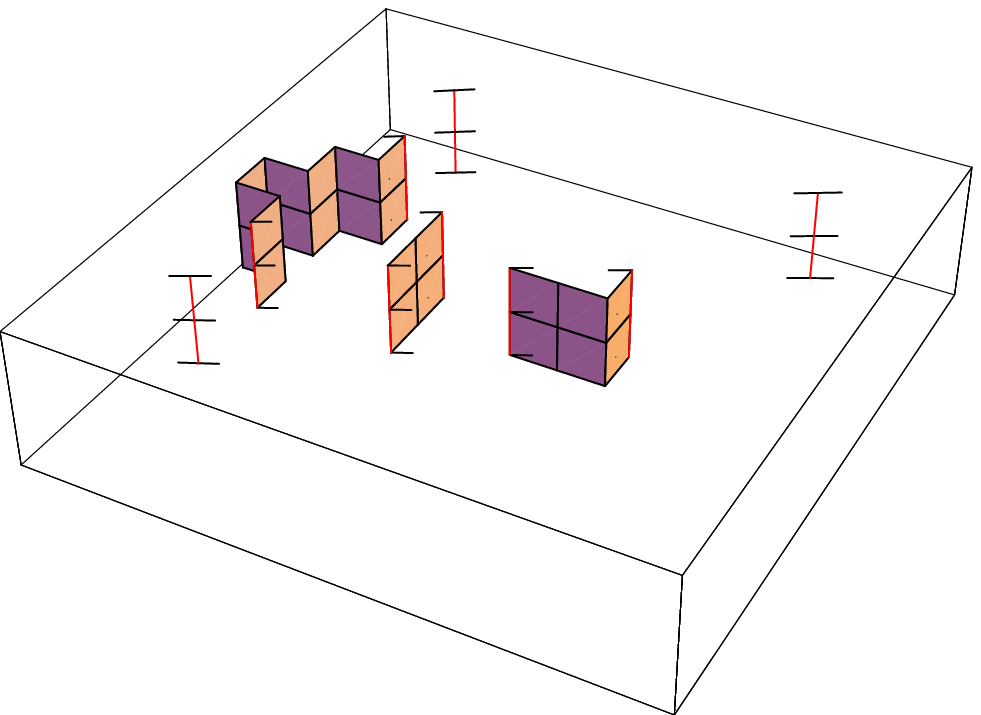}}}
\hfill
\subfigure[\hfill]{\scalebox{0.60}{\includegraphics{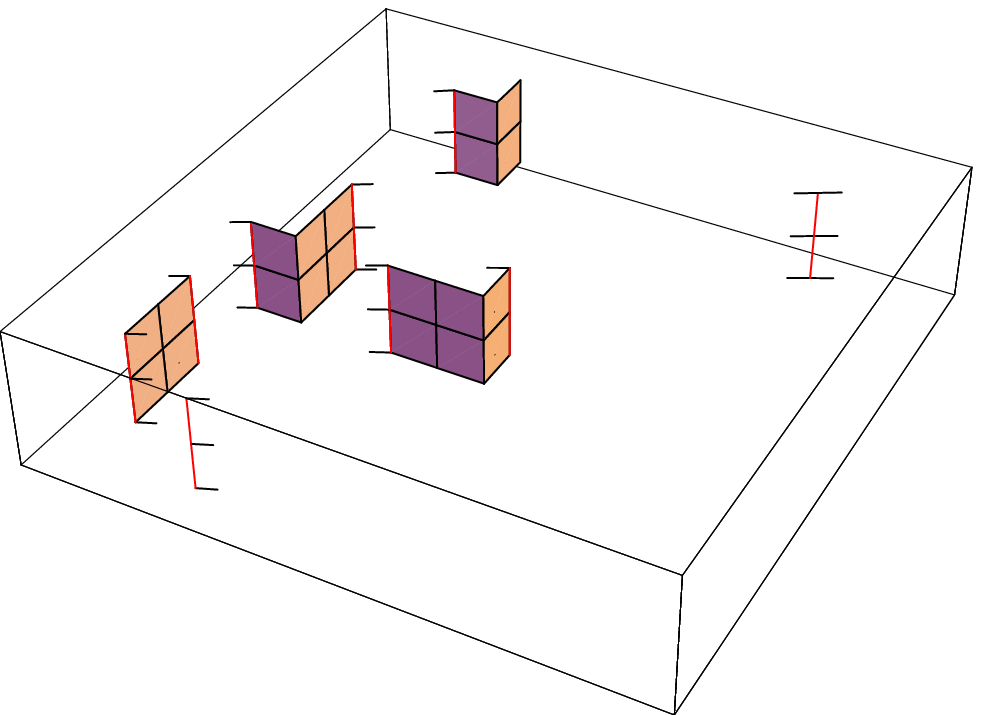}}}
\hspace{5mm}
\caption{Dual P-plaquettes in a typical field configuration at 
$\beta=2.6$, on a $2\cdot 12^3$-lattice. Two successive z-slices for 
the
x-y-t-subspace are shown. The amputated lines leaving the left figure 
towards
right arrive in the right figure from the left.}
\label{mathft}
\end{figure}
\begin{figure}
\centering
\centerline{\scalebox{0.8}{\includegraphics{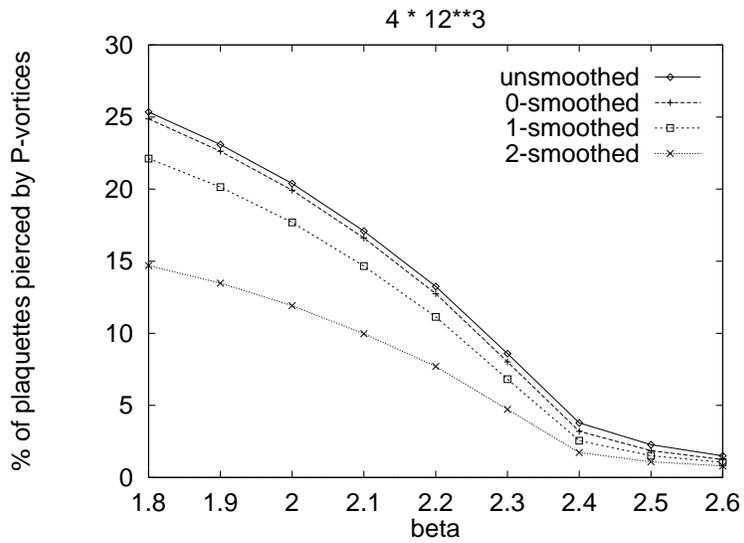}}}
\caption{Percentage of P-plaquettes as a function of $\beta$ for a 
$4\cdot
12^3$-lattice.}
\label{ftdens}
\end{figure}
Fig.~\ref{mathft} displays the dual P-plaquettes of a typical 
field configuration at $\beta=2.6$ 
on a $2\cdot 12^3$-lattice. The dual P-plaquettes form cylinders in 
time
direction, closed via the periodicity of the lattice. Vortices of this 
shape
are also well known in finite temperature theory under the name of 
ordered-ordered interfaces \cite{HJK92}.

The density of P-plaquettes is depicted in Fig.~\ref{ftdens} 
for various smoothing steps. The decrease in the number of P-plaquettes 
in the
0-smoothing step is almost independent of $\beta$. The number of 
P-plaquettes
above the phase transition decreases rather slowly. 

\begin{figure}
\centering
\centerline{\scalebox{0.8}{\includegraphics{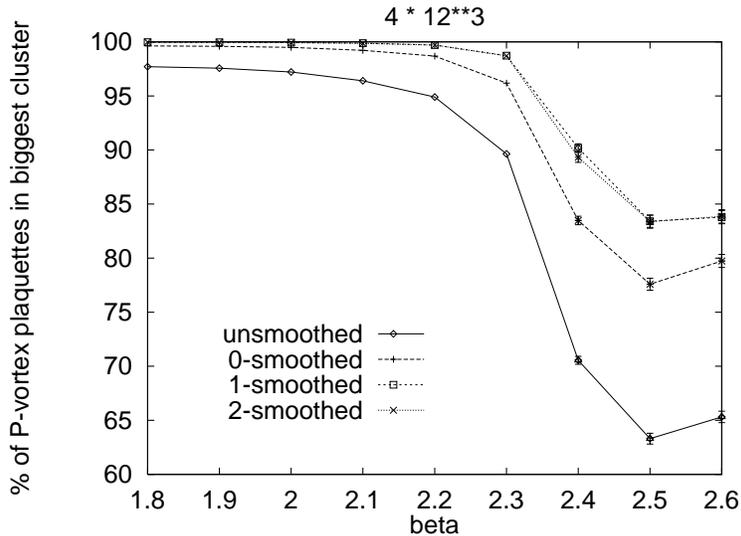}}}
\caption{Size of largest vortex in \% of all P-plaquettes for a $4\cdot 
12^3$-lattice.}
\label{ftBGCL}
\end{figure}

As expected from the zero temperature results in the confined phase 
most of
the P-plaquettes belong to a single large vortex. This can be seen in 
Fig.~\ref{ftBGCL} .The situation changes drastically at the phase 
transition
where the percentage of P-plaquettes in the largest vortex drops 
considerably, especially for the unsmoothed configurations. The 
smoothing
procedure shows that there is still one largest vortex but its 
dominance is
not so strong as in the zero-temperature case. The increase of the 
percentage
from $\beta = 2.5$ to $\beta = 2.6$ could be just a finite size effect;
this will require investigation on larger lattices.
In any case, the existence of a large space-time vortex on the dual
lattice is required, at finite temperature in the deconfinement phase,
in order to explain area 
law behaviour for spacelike Wilson loops.

\begin{figure}
\centering
\centerline{\scalebox{0.8}{\includegraphics{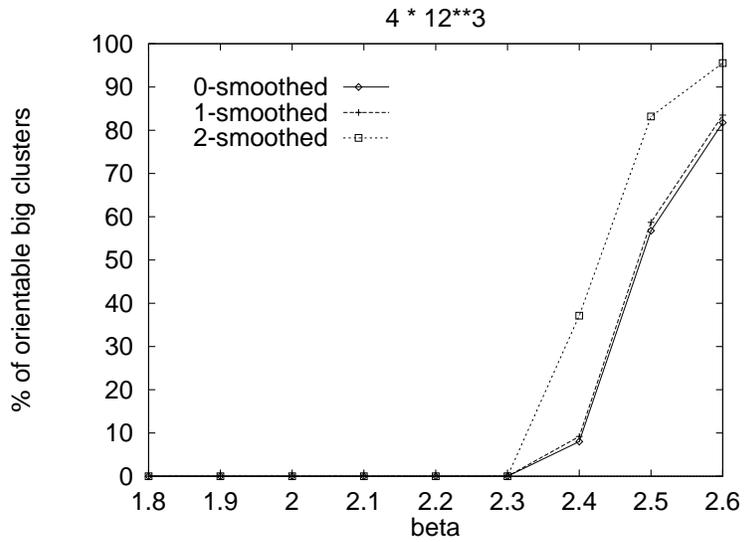}}}
\caption{Percentage of orientable P-vortices for a $4\cdot 
12^3$-lattice.}
\label{ftorient}
\end{figure}
With the decrease in the percentage of E-plaquettes we find increasing 
orientability of P-vortices in Fig.~\ref{ftorient}. The orientability 
approaches 100\% for large $\beta$. The smoothing procedure shows which 
part
of the unorientability is due to short range fluctuations of the vortex.

\begin{figure}
\centering
\centerline{\scalebox{0.8}{\includegraphics{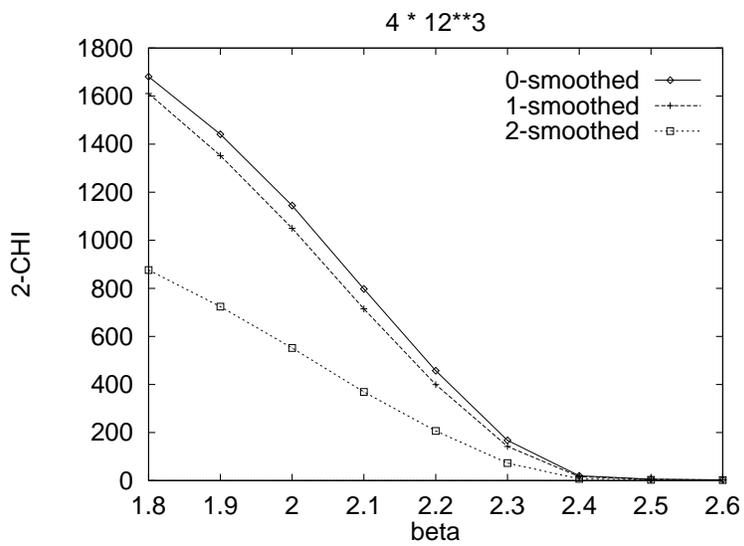}}}
\caption{Genus of largest vortex for a $4\cdot 12^3$-lattice.}
\label{ftgenus}
\end{figure}

The relations between genus $g$ and Euler characteristic $\chi$ are 
different
for orientable (\ref{Euler-orient}) and unorientable surfaces 
(\ref{Euler-unorient}). Since both types of surfaces appear in finite 
temperature calculations we investigate the value of $2-\chi$ as in the 
zero-temperature case. For orientable vortices this expression is the 
genus
$g$, for unorientable surfaces half of the genus. In Fig.~\ref{ftgenus} 
we
show the value of $2-\chi$ for the largest vortex. These data are not 
scaled
with the lattice constant $a$ as in the zero-temperature case 
(Fig.~\ref{euler}). In the confined phase P-vortices are again 
complicated
surfaces  and especially 2-smoothing reduces the number of handles. 
Above the
phase transition $2-\chi$ approaches the value~2. 
This is a consequence of the 
vanishing density of E-plaquettes. The largest P-vortex becomes 
orientable
with genus $g=1$ and $\chi=0$. It has the topology of a torus.

\section{Conclusions}

We have investigated the size and topology of P-vortices in SU(2) 
lattice gauge theory; P-vortices are surfaces in the dual lattice which 
lie at or near the middle of thick center vortices.  We have found that in 
the confined phase the four-dimensional lattice is penetrated by a single 
huge P-vortex (see also \cite{Po98}) of very complicated topology.  
This huge P-vortex is a closed surface on the dual lattice which 
is unorientable and has many ($\sim 10/{\rm fm}^4$) 
handles. There exist also a few very small vortices. The short range 
fluctuations of the large P-vortex contribute only to the perimeter law
falloff of projected
Wilson loops.  These short range fluctuations may simply be due to a slight
ambiguity in the precise location of the middle of a thick center vortex, as 
was discussed in section 4 of ref.\ \cite{PRD98b}, and are not necessarily
characteristic of the thick center vortices themselves.  

  By a smoothing procedure, we were able to remove these perimeter 
contributions due to short-range fluctuations, keeping the Creutz ratios 
constant. Thus the short-range P-vortex fluctuations are found to account for 
the difference between the percentage $p$ of plaquettes which are pierced 
by P-vortices, and the comparatively smaller fraction $f$ which, in the
simplest version of the vortex model (with uncorrelated P-plaquettes),
contribute to the string tension.  Upon smoothing away the short-range
fluctuations, we find the fraction $p$ closely approaching the value of
$f$ extracted from the asymptotic string tension.

  The density of vortices does not vanish in the deconfined phase, but 
there is found to be a strong space-time asymmetry.
P-vortices at finite temperature are mainly 
composed of space-space plaquettes forming \emph{timelike} surfaces 
on the dual lattice.  These surfaces are closed via the periodicity of
the lattice in the time direction, they are orientable,
and have the topology of a torus, i.e. genus 
$g=1$. The dominance of the largest vortex is not as strong as in the 
zero temperature case.  The space-time asymmetry of P-vortices in the 
deconfined phase nicely explains \cite{LT98} the corresponding asymmetry 
in Wilson loops, which have area-law falloff for spacelike, and vanishing 
string tension for timelike loops.   

\bigskip
\noindent {\Large \bf Acknowledgements}
\bigskip

This work was supported in part by Fonds zur F\"orderung der
Wissenschaftlichen Forschung P11387-PHY (M.F.), 
the U.S. Department of Energy under Grant No. DE-FG03-92ER40711
(J.G.), the ``Action Austria-Slovak Republic: Cooperation
in Science and Education'' (Project No. 18s41) and the Slovak Grant 
Agency
for Science, Grant No. 2/4111/97 (\v{S}.O.).

\end{document}